# Common-view mode synchronization as a source of error in measurement of time of flight of neutrinos at the CERN-LNGS experiment

Satish Ramakrishna
Nepean Sea LLC, 296 Harlingen Road, Belle Mead, NJ 08502

The CERN-LNGS time-of-flight (TOF) experiment (of neutrinos) represents a significant challenge to the special theory of relativity and needs to be addressed either as a source of new physics, or as an un-remedied experimental error. There have been several attempts at using new physics to explain the results, while a few unpublished results exist that address the experimental errors that might lead to the same result. In particular, a recent calculation by van Elburg indicates a potential flaw in the OPERA experiment that represents a source of potential error, i.e., that the motion of the synchronizing GPS satellites causes the clocks to go out of synchronization. Unfortunately, there are several misconceptions about how GPS satellites work that pervade the paper. In addition, the principal contention of the paper, that the experiment is being timed by a moving clock is not substantiated in the analysis. This paper substantiates the point, as well as introduces a new source of error in the "common-view" method of synchronization of clocks.

The OPERA experiment is conceptually simple – neutrinos are emitted (with a distribution in emission times) from the CERN emitter, detected at LNGS, under Gran Sasso (again, distributed in reception times). With the clocks at both locations being properly synchronized, the travel time is calculated. The experiment finds that the neutrinos are received at LNGS $60 \pm 6.9$ (stat.) $\pm 7.4$ (sys.) nanoseconds (ns) earlier than light would have been received. This indicates that neutrinos have travelled faster than light. Several unpublished explanations have been adduced [5...10]. The calculation by van Elburg [2] represents an interesting partial solution to the puzzle.

The author of [2] points to apparently relativistically uncorrected clocks on the GPS satellites as being the source of the problem. In fact, the general relativistic effect (clocks slowing down nearer large masses, so the Earth clock would run 45 μs per day <u>slower</u> than the GPS satellite's clock) is larger than the special relativistic effect (which would make the Earth clock run 7 μs per day <u>faster</u> than the GPS satellite's clock). The clocks on the GPS satellites are themselves corrected for this effect – were they not, they would run 38 μs per day <u>faster</u> than the corresponding clocks on the Earth. If this were allowed to persist, once initially calibrated, planes would drift 10.8 kilometers from their prescribed places after one day (since time of flight is used to calculate distances from these satellites to obtain navigational fixes). To prevent this from happening, the clocks are physically engineered, prior to launch into orbit, to run 38 μs per day slower than a regular clock, so they continually compensate for the difference.

The point made in [2] is that the GPS satellite (that is used to calibrate the clocks at the source and destination) itself is moving along the direction of the beam and this accounts for a shorter time-of-flight measured on board the satellite clock. A proper calculation would lead to a difference between the OPERA collaboration's calculation and the correct calculation for the time of flight between the emitter and detector. Using the relevant numbers for the set-up (described in the last section), this calculation leads to a difference of 28 nanoseconds in the time of travel, which is about 45% of the observed time difference [1]. However, the fact that the clocks used in the experiment are synchronized to GPS satellites is not sufficient evidence that the experiment is being timed in the satellite reference frame, since the propagation delays are apparently being taken into account in the synchronization.

**Error due to Common View Method of Synchronization**

The method of synchronization of clocks in the CERN-LNGS experiment is described very precisely in the references [11...17]. The basic approach of this "common-view mode", described in [11] is shown in Figure 1. The same GPS satellite is observed at the same MJDs (time range in Modified Julian Date format) from the two stations A (CERN) and B (LNGS). The GPS satellite sends out its position as well as the time-stamp from its internal atomic clock to both the stations A and B. At each station, the GPS clock time-stamp is precisely noted along with the local time according to the local GPS-disciplined atomic clock. We assume, along the lines of [2], that the GPS satellite is moving along the A-B axis, towards B, which is consistent with the satellite's orbit, on a plane inclined at 55° to the equatorial plane and from west to east.

Fig. 1: Common-view mode synchronization

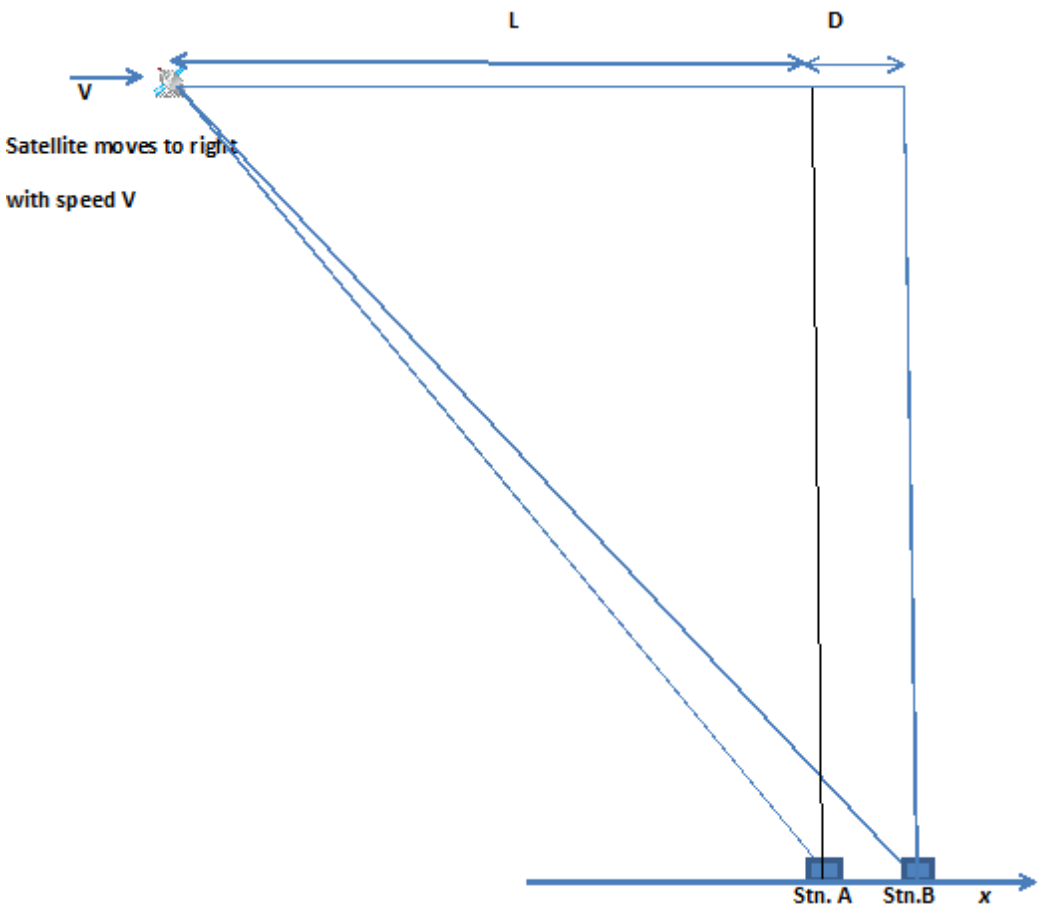

If $T_A$ and $T_B$ are the local time stamps both corresponding to the same GPS time stamp $T_{GPS}$, then we can relate these quantities by

$$\begin{cases} T_A = T_{GPS} + PD_{sat-A} + \delta_A \\ T_B = T_{GPS} + PD_{sat-B} + \delta_B \end{cases} \quad (1)$$

where $PD_{sat-A}$ and $PD_{sat-B}$ are, respectively, the propagation delays from the satellite to A and B (measured by clocks in the earth frame) and $\delta_A$ and $\delta_b$, respectively, the delays between the two clocks at A and B on the earth and this GPS time. However, this is not how the clocks are set. They are actually set to

$$\begin{cases} T_A = T_{GPS} + PD'_{sat-A} + \delta_A \\ T_B = T_{GPS} + PD'_{sat-B} + \delta_B \end{cases} \quad (2)$$

Here $PD'_{sat-A}$ and $PD'_{sat-B}$ are, respectively, the propagation delays from the satellite to A and B (measured by the GPS-disciplined clocks and the GPS satellite's own clock, hence the tilde – they are measured in the satellite's reference frame). Then, to eliminate the GPS time from the measurement, the common-view method considers the difference

$$T_B - T_A = (PD'_{sat-B} - PD'_{sat-A}) + (\delta_B - \delta_A) \quad (3)$$

In the CERN-LNGS experiment, much effort was expended on ensuring that the number $(\delta_B - \delta_A)$ was as small as possible [1]. In particular, independent measurements by the PTB (Physikalische-Technische Bundesansalt) [16] and the Swiss Metrology Institute (METAS) [17] show that this difference is under 3 ns.

These propagation delays need to be considered carefully between the two frames. Since they are computed using the GPS clock on the satellite and the GPS-disciplined clocks (which are feedback-controlled to follow the GPS clocks with short- and long-term stability [18]), they need to be relativistically corrected. In particular, using the notation in Figure 1,

$$PD'_{sat-A} = \frac{1}{\sqrt{1-\frac{V^2}{c^2}}} \left( PD_{sat-A} - \frac{V L}{c^2} \right) \quad (4)$$

$$PD'_{sat-B} = \frac{1}{\sqrt{1-\frac{V^2}{c^2}}} \left( PD_{sat-B} - \frac{V(L+D)}{c^2} \right) \quad (5)$$

We find

$$PD'_{sat-B} - PD'_{sat-A} = \frac{1}{\sqrt{1-\frac{V^2}{c^2}}} \left( PD_{sat-B} - PD_{sat-A} - \frac{V D}{c^2} \right) \quad (6)$$

Note that $V^2/c^2$ is very small and the term $1/\sqrt{1-\frac{V^2}{c^2}}$ can be safely ignored in the rest of the analysis – it is too close to unity to matter for the conditions considered here. As a consequence

$$PD_{sat-B} - PD_{sat-A} \approx (PD'_{sat-B} - PD'_{sat-A}) + \frac{V D}{c^2} \quad (7)$$

i.e., the propagation delay difference (between stations B and A) in the earth frame (the "real" propagation delay) is longer by the quantity $V D/c^2$. This in turn means that the clock at B is being set $V D/c^2$ earlier than it ought to be, even after synchronization using the methods employed in the experiment. The quantity $V D/c^2$ is roughly 28 ns, as calculated in the last section, so the TOF experiment would indicate that the neutrinos were being received earlier than they were expected to, after their departure from A.

One method of avoiding this problem would be to synchronize the clocks with the actual, earth-frame propagation delay. This would indeed imply that we use (1) to set the clocks, though, as shown in the next section, since the propagation delays in the earth frame are basically identical, this simply implies that the clocks are just being set to and forced to precisely follow the GPS satellite clock. This then implies that the clocks are being synchronized in the GPS satellite's reference frame, up to $(\delta_B - \delta_A)$. If two events were to occur, the first at Station A and then a second one at Station B, with a time interval $\Delta t$ (in the earth frame) and an interval $\Delta t'$ in the satellite frame, the relationship between these time intervals is

$$\Delta t' = \frac{1}{\sqrt{1-\frac{V^2}{c^2}}}\left(\Delta t - \frac{V D}{c^2}\right) \tag{8}$$

i.e., if the two events were synchronized in the satellite frame (the clocks being set to the same time), $\Delta t' = 0$, which implies $\Delta t = \frac{V D}{c^2}$. This means that the events are not simultaneous in the earth frame and additionally, the clock at B is being set early compared to where it ought to be set. Again the clock is set $V D/c^2$ too early, which would reduce the measured flight time by 28 ns.

**Error due to Time measurement in the GPS framework**

The clocks in the experiment are both run with very controlled values of $(\delta_B - \delta_A)$ under 3 ns. The times at the clocks at the stations A and B can therefore be written as expressions similar to equations (1) and (2). Next, we consider the various events involved in the experiment – Event #1 is the emission of neutrinos at the CERN synchrotron (station A) and Event #2 is the reception of neutrinos at the LNGS site (station B). The following table (Table 1) is useful (we assume that the difference $(\delta_B - \delta_A)$ is 0.

In the table the notation A:2 and B:2 refer to the instants on the GPS clock (not simultaneous in the GPS frame) when the event (2) occurs at B and is assumed to be timed simultaneously at A. Also, based on our analysis in the previous section we will use the "real", i.e., earth frame propagation delays.

The time of flight of the neutrinos is basically

$$T_B(2) - T_A(1) = \left(T_{GPS}(B:2) + PD_{sat-B}(B:2)\right) - \left(T_{GPS}(1) + PD_{sat-A}(1)\right)$$

$$= \left(T_{GPS}(B:2) - T_{GPS}(1)\right) + \left(PD_{sat-B}(B:2) - PD_{sat-A}(1)\right) \tag{8}$$

Table 1: Events and their times

| TIME | EVENT #1: Emission of neutrinos at CERN – Station A | EVENT #2: Reception of neutrinos at LNGS – Station B |
|---|---|---|
| **Atomic clock on GPS satellite** | $T_{GPS}(1)$ | $T_{GPS}(2)$ |
| **Clock at A** | $T_A(1) = T_{GPS}(1) + PD_{sat-A}(1)$ | $T_A(2) = T_{GPS}(A:2) + PD_{sat-A}(A:2)$ |
| **Clock at B** | $T_B(1) = T_{GPS}(1) + PD_{sat-B}(1)$ | $T_B(2) = T_{GPS}(B:2) + PD_{sat-B}(B:2)$ |

Let's analyze the quantity

$$(PD_{sat-B}(B:2) - PD_{sat-A}(1))$$

Consider the simplified situation, in Figure 2, where the satellite is at a height $H$, midway between the two stations at the point when it sends a signal that triggers event (1) and then travels to the right (towards station B), at speed $V$, in the time interval $\Delta t$ when Event #2 occurs at B.

Fig. 2: Simplified description of propagation delays for TOF measurement

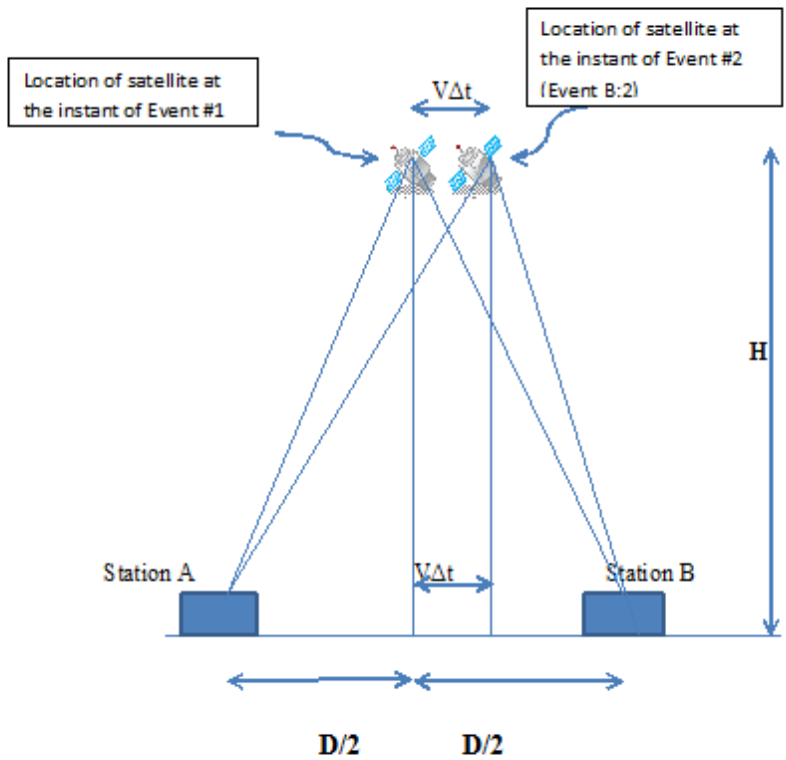

Basic geometry implies,

$$PD_{sat-A}(1) = \frac{\sqrt{H^2 + \frac{D^2}{4}}}{c}, \quad PD_{sat-B}(B:2) = \frac{\sqrt{H^2 + \frac{\left(\frac{D}{2} - V\Delta t\right)^2}{4}}}{c}$$

so that, to first order in $V \Delta t$,

$$(PD_{sat-B}(B:2) - PD_{sat-A}(1)) = -\frac{1}{2}\frac{V \delta t}{c}\frac{D}{\sqrt{H^2 + \frac{D^2}{4}}} \qquad (9)$$

Which is a factor 1/30 times smaller than the other errors computed in this paper, since **D** is roughly 730 km, while **H** is roughly 20,200 km. For a time interval corresponding to flight of photons from A to B, this difference is approximately 0.55 ns.

For all practical purposes, we can write

$$T_B(2) - T_A(1) \approx T_{GPS}(B:2) - T_{GPS}(1) \quad (10)$$

The point being made here is that the GPS satellite is so far away that the propagation delays are negligibly different to the various points on the earth used in the experiment. This fact helps remove differences due to ionospheric/tropospheric variations between the two stations, the result is that the earth clocks are basically synchronizing to a moving clock continually – this is especially true for GPS-disciplined clocks.

Now we can deduce the result in paper [2] accurately. Let's assume that we have three clocks in the situation. One is a clock on the Earth, the second is travelling on the GPS satellite and is uncorrected for any relativistic effects (special and general) in its construction and the third is also on the GPS satellite, but also corrected for all relativistic effects. The two Events #1 and #2 are, as before,
- emission of the neutrino, represented by $(x_1, t_1)$ in the earth frame and $(x_1', t_1')$ in the satellite's frame, time being measured in this frame with clock #2 and
- reception of the neutrino, represented by $(x_2, t_2)$ in the earth frame and $(x_2', t_2')$ in the satellite's frame, time being measured in this frame again with clock #2.

The time of flight measured by the clock on the ground should be

$$\tau = t_2 - t_1 = \frac{x_2 - x_1}{c} = \frac{D}{c} \quad (11)$$

The time of flight measured by the uncorrected (second) clock on the satellite travelling along the CERN-Gran Sasso line (towards the detector at Gran Sasso) at the speed $V$ is

$$\tau_{clock2} = t_2' - t_1' = \frac{1}{\sqrt{1-\frac{V^2}{c^2}}} \left( \tau - \frac{V}{c^2} D \right) \quad (12)$$

which accounts for the Lorentz contraction [4] along the direction of travel of the satellite, as well as for the fact that from the point of view of this clock, the detector is approaching the neutrinos at the speed $V$.

The actual time-interval measured by the corrected clock (#3) is

$$\tau_{clock3} = (1 + f) \tau_{clock2} \quad (13)$$

Where $f$ is roughly 38 µs per day (i.e., every 86,400 seconds). This is roughly $4.28 \times 10^{-10}$, so that for all practical purposes $\tau_{clock3}$ and $\tau_{clock2}$ are equal. Note that the relevant point is that, in the satellite's reference frame, we need to consider <u>the detectors' motion towards the neutrinos</u> as well as the less significant Lorentz contraction of the baseline length along the direction of the satellite's motion.

If the time measurements used in this experiment involved averaging over the GPS signals received from several satellites at once, this effect would probably average out, however, since the possible orbits of the GPS satellites are either roughly parallel to the CERN-LNGS line or roughly perpendicular to it, the averaging will, at most, reduce the size of the effect not cancel it out (since it is the component of GPS satellites' velocity along the CERN-LNGS line that matters). The details would depend on how many satellites were averaged over and what their precise speeds and directions of travel were - the transit time for the neutrino flight is barely 2.4 milliseconds.

**Calculation**

We proceed to a numerical calculation. The orbital speed of the satellite is (at a distance of 26,200 km away from the earth's center) is roughly 3810 km/s. The earth is also spinning while this orbit is taking place, at a speed of 328 m/s at the 45° latitude of Milan (roughly half way between the detector and the emitter), whose component along the 55° angle to the latitudes of the satellite's orbit is 232 m/s. This results in a time-difference (note expression (13)) of 28 ns between the earth-frame time difference $\tau$ and the satellite-frame time-difference $\tau_{clock3}$, note that $\tau_{clock3} < \tau$. The other part of the difference comes from equation (7), which indicates that the clock at station B is slightly early, by another 28 ns. The total difference is the sum of the two, i.e., the neutrinos appear to arrive 56 ns <u>earlier</u> than expected by the speed of light.

There is a simple way to check the statements made here – they involve doing the same experiment with light. The results would be identical to the results obtained with neutrinos, i.e., photons would appear 60 ns before they were supposed to. The clocks need to be synchronized identically of course and the experiment could be done over the ground.

**Conclusion**

Our calculation above is a corrected version of van Elburg's idea and explains why there is an approximately 60 ns difference in the arrival time of neutrinos and what is expected based on the speed of light. The problem lies in the method used to synchronize and set the clocks – there is nothing specifically wrong with using GPS atomic clocks, even for synchronization, except that the clocks should not be set to follow the GPS clock after the point of synchronization. This is unfortunately what happens with the PolaRx2e receivers used as the time-bases in the set-up. If, instead, the Cs4000 clock were the only clocks used, this result would not appear.

The author would like to thank Dr. K. Aiyer for useful comments.